\documentclass[aps,pra,showpacs,amsfonts,amssymb,amsmath,twocolumn]{revtex4} 
\usepackage{times}
\usepackage{epsfig}

\newcommand{\MP}{}
\newcommand{\SP}{\mathcal}

\begin{document}

\title{Quantum theory of feedback of bosonic gases}

\author{S. Wallentowitz}

\email{sascha.wallentowitz@physik.uni-rostock.de}

\affiliation{Fachbereich Physik, Universit\"at Rostock,
  Universit\"atsplatz 3, D-18051 Rostock, Germany}

\date{March 13, 2002}

\begin{abstract}
  A quantum theory of feedback of bosonic many-atom systems is
  formulated. The feedback-induced many-atom correlations are treated
  by use of a parameterized correlation function, for which closed
  equations of motion are derived. Therefrom the dynamics of any
  additive property of the system, i.e., properties derived from the
  reduced single-atom density operator, can be obtained. An example is
  given that indicates the correlation effects of feedback.
\end{abstract}


% 03.65.-w  Quantum mechanics, field theories, and special relativity
% 03.65.Ca  Formalism
% 39.25+k   Atom manipulation (scanning probe microscopy, laser
%           cooling, etc.)  
% 05.30.-d  Quantum statistical mechanics
% 42.50.Lc  Quantum fluctuations, quantum noise, and quantum jumps
% 45.80.+r  Control of mechanical systems
% 03.67.-a  Quantum information
% 03.75.Fi  Phase coherent atomic ensembles; quantum condensation
%           phenomena
% 05.30.Jp  Boson systems (for Bose-Einstein condensation, see
%           03.75.Fi)

\maketitle

\section{Introduction} \label{sec:1}

Feedback is widely used in engineering to control certain properties
of a system.  Motivated by optical homodyne measurements, a theory of
feedback has been formulated for cavity fields~\cite{wiseman}.  Many
applications of feedback have been proposed for single quantum
systems, such as, besides single-mode cavity fields, also for single
atoms or generic harmonic oscillators.  Moreover, extensions to
time-delayed non-Markovian feedback have been
derived~\cite{non-markovian-fb} and coherent
feedback~\cite{lloyd-coherent} and feedback using weak
measurements~\cite{lloyd-weak} have been studied.  Furthermore,
alternative approaches to feedback have been presented in the context
of quantum-state estimation and quantum control~\cite{alternative-fb}.

Besides laser cooling~\cite{cooling} many advances have been
accomplished in the last years in cooling atoms to the ultimate
limits. Among them are Bose--Einstein condensation of dilute gases of
alkali species~\cite{bec-alkali1,bec-alkali2,bec-alkali3,bec-alkali4}
and atomic hydrogen~\cite{bec-hydrogen}, the experimental realization
of atom lasers~\cite{atom-laser1,atom-laser2,atom-laser3} and the
implementation of micro- and nanostructured atom-optical
devices~\cite{bec-microstructures1,bec-microstructures2,bec-microstructures3}.
This experimental progress draws attention to the investigation of
many-atom systems and their condensate features from a quantum-optical
viewpoint. In view of the experimental capabilities up to date, it is
quite promising to engage in developing methods for controlling
complex many-atom systems, such as produced nowadays in the
laboratory.

The use of feedback methods acting on accessible macroscopic
observables of the system appears particularly desirable. An
experimental implementation of it would not require the active
intervention of the experimenter, but would act as a self-automated
basic technological tool, such as it is now in engineering.
Unfortunately, for the case of feedback acting on a multi-mode
many-atom system, the existing descriptions of feedback are in general
not applicable: When trying to access arbitrary properties of the
system, these methods become intractable, due to the required vast
number of degrees of freedom.

Typically employed mean-field approximations, could be used to
factorize or truncate higher-order many-atom correlations, in order to
simplify the problem.  In the considered case, however, the feedback
operation contains the measurement of a macrostate observable of the
many-atom system, to which all atoms contribute. The collective access
to the atoms in such a measurement induces strong correlations between
them.  These correlations represent a main ingredient of the feedback
and should be taken care of in a non-approximate way. The formation of
correlations becomes clear when considering that an observed
measurement outcome of a macrostate observable can be produced by a
vast number of different microstates, that all enter the feedback
operation in a coherent superposition.  Such superposition states may
contain strong correlations between the atoms and one may therefore
expect that in general many-atom quantum correlations play a dominant
role in the dynamics of the feedbacked system.

Correlations of this type are usually generated by interactions
between atoms, such as for example two-body collisions. In the case
considered here, such explicit interactions will not be taken into
account.  The many-atom correlations will solely be generated by the
measurement process. An interaction between the atoms can in fact be
identified in the measurement.  The measurement process can be thought
of as a probe field interacting with all the atoms with a subsequent
projection of the probe by the final readout of the measurement
result. Thus we may imagine an effective interaction between the atoms
being mediated via the probe field. 

In the following an exact method is presented for describing the
reduced single-atom dynamics under the influence of feedback
operations acting on macroscopic observables of a bosonic multi-mode
many-atom system.  The feedback loop consists of a measurement with
predefined measurement uncertainty $\sigma$ and a subsequent unitary
operation that again acts on all atoms of the system.  The feedback
loop is assumed to act instantaneously without time delay between
measurement and operation. As will be shown, the many-atom
correlations, that are generated by the measurement, can be formally
incorporated into the single-atom dynamics by introducing a
dynamically evolving additional degree of freedom. Thus the free
evolution intermitted by feedback operations can be formulated in
terms of a quasi single-atom dynamics, revealing modifications to the
feedback dynamics of a true single-atom system.

The paper is structured as follows: In Sec.~\ref{sec:2} the
macroscopic observables needed for the feedback are introduced.
Furthermore, the effect of the feedback on the many-atom quantum state
is described and a first indication of the importance of atom-atom
correlations is given. Section~\ref{sec:3} proceeds by introducing a
specially designed correlation function that is shown to allow for a
closed description of the feedback. Besides the dynamical evolution
during the free evolution between feedback operations, also an
interpretation of this correlation function is presented. An
application of the developed method to the case of a condensed bosonic
gas is given in Sec.~\ref{sec:4} and finally a summary and conclusions
are found in Sec.~\ref{sec:5}.

\section{Feedback loop} \label{sec:2}
 
\begin{figure}
  \begin{center}
    \epsfig{file=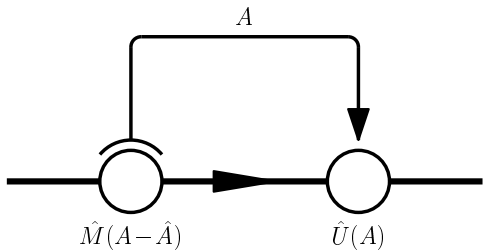,scale=0.55}
    \caption{Schematic function of the feedback loop. First a
      measurement on the system occurs, described by the operator
      $\hat{M}(A \!-\!  \hat{A})$. The value of the measured outcome
      $A$ is then used as a parameter for the unitary operation
      $\hat{U}(A)$.}
    \label{fig:scheme}
  \end{center}
\end{figure}

The feedback loop shall be considered as a measurement of a
macroscopic observable $\hat{A}$ of the system and the use of the so
obtained information, i.e., the measured value $A$, for applying a
unitary operation, $\hat{U}(A)$, on the system. Information is
therefore gained by measurements and used to manipulate the system,
cf. Fig.~\ref{fig:scheme}.  
For the macroscopic observable we may think of any extensive quantity
of the system, i.e., a quantity that has additive contributions from
each atom in the system. Thus measurements are not performed on single
atoms but on large numbers of atoms.

First let us consider the single-atom observable $\hat{\SP A}$ and its
canonical conjugate $\hat{\SP B}$ with commutator ($\hbar \!=\! 1$)
\begin{equation}
  \label{eq:sp-comm}
  [\hat{\SP A}, \hat{\SP B}] = i .
\end{equation}
Note that from now on for denoting single-atom Schr\"odinger operators
calligraphic symbols are used.  We now turn to a second-quantized
description of the many-atom system by introducing the bosonic
atom-field operator $\hat{\phi}(x)$ with commutator
\begin{equation}
  [\hat{\phi}(x), \hat{\phi}^\dagger(x')] = \delta (x \!-\! x') .
\end{equation}
Using an expansion into a set of complete and orthonormal modes
$u_\lambda(x)$,
\begin{equation}
  \hat{\phi}(x) =  \sum_\lambda \hat{\phi}_\lambda
  u_\lambda(x) , 
\end{equation}
the resulting commutator relation for the annihilation and creation
operators of atoms in mode $\lambda$, $\hat{\phi}_\lambda$ and
$\hat{\phi}_\lambda^\dagger$, respectively, reads
\begin{equation} 
  [\hat{\phi}_\lambda, \hat{\phi}_\mu^\dagger ] = \delta_{\mu\lambda}
  . 
\end{equation}

The macroscopic observable $\hat{A}$ that is used for the measurement
in the feedback loop is a sum of contributions of the individual
atoms. In the formalism of second quantization $\hat{A}$ is therefore
bilinear in the atom-field operators and reads
\begin{equation}
  \label{eq:ops-mp}
  \hat{A} = \sum_{\mu\lambda} 
  \hat{\phi}_\mu^\dagger \hat{\phi}_\lambda
  \, \langle \mu | \hat{\SP A} | \lambda \rangle .
\end{equation}
Here $\langle \mu | \hat{\SP A} | \lambda \rangle$ are the matrix
elements of the corresponding single-atom operator $\hat{\SP A}$ with
the single-atom Schr\"odinger modes $|\lambda\rangle$ being defined
by: $\langle x | \lambda \rangle \!=\!  u_\lambda(x)$.

For implementing a feedback operation $\hat{U}(A)$ that is capable of
changing the measured property $A$, the canonically conjugate variable
of $\hat{A}$ is needed. Whereas for a single atom this is trivially
given by the operator $\hat{\SP B}$, cf.~Eq.~(\ref{eq:sp-comm}), for a
general many-atom system a conceptual problem arises. Clearly, the
proper canonical conjugate is given by
\begin{equation}
  \label{eq:mp-proper-B}
  \hat{B} = \frac{1}{\hat{N}} \sum_{\mu\lambda} \hat{\phi}_\mu^\dagger
  \hat{\phi}_\lambda \langle \mu | \hat{\SP B} | \lambda \rangle ,
\end{equation}
where again $\langle \mu | \hat{\SP B} | \lambda \rangle$ are the
matrix elements of the corresponding single-atom operator $\hat{\SP
  B}$. Moreover, $\hat{N}$ is the operator of the total number of
atoms in the system,
\begin{equation}
  \label{eq:2}
  \hat{N} = \sum_\lambda \hat{\phi}_\lambda^\dagger \hat{\phi}_\lambda .
\end{equation}
With this definition the sought commutator relation is established:
\begin{equation}
  \label{eq:mp-proper-comm}
  [\hat{A}, \hat{B} ] = i , 
\end{equation}
as in the case of a single atom, cf.~Eq.~(\ref{eq:sp-comm}).

However, due to the appearance of the atom-number operator $\hat{N}$
in Eq.~(\ref{eq:mp-proper-B}), the operator $\hat{B}$ is not
bidiagonal in the atom-field operators. It is an intensive quantity
and not an extensive one and thus is not a simple sum of contributions
of the individual atoms. Thus it cannot be experimentally realized by
application of an external field (e.g. laser field) that interacts
with each individual atom. In other words, the problem is, that the
total number of atoms is in principle unknown without its particular
measurement. For the feedback action we need to know how many atoms
have contributed to the previously measured quantity $A$. The lack of
this knowledge is a potential source of imperfection of the overall
feedback action.

This problem can be overcome in two ways: Either in each feedback loop
not only $\hat{A}$ but also the atom number $\hat{N}$ is measured, or
the atom number is estimated. We will here concentrate on the latter
approach, where the estimated number of atoms $N_{\rm e}$ may be
derived from a priori knowledge about the system, without additional
measurements.

It is worth noting, that neglecting this issue by replacing $\hat{N}$
by its average $\langle \hat{N} \rangle$ in
Eq.~(\ref{eq:mp-proper-B}), but still assuming the commutator
relation~(\ref{eq:mp-proper-comm}), results in the neglect of
atom-number fluctuations. Such an approach is only valid when the atom
number is exact. Considering the case where only part of the system is
subject to the feedback, say only atoms being located in a predefined
spatial region, an exact atom number in this region is rather doubtful
even when the total number is precisely known.

Such an approximation can be avoided by consistently defining the
operator $\hat{B}$ by use of the estimated atom number $N_{\rm e}$,
\begin{equation}
  \label{eq:ops-mp2}
  \hat{B} = \frac{1}{N_{\rm e}} \sum_{\mu\lambda} 
  \hat{\phi}_\mu^\dagger \hat{\phi}_\lambda
  \, \langle \mu | \hat{\SP B} | \lambda \rangle ,
\end{equation}
and using the resulting commutator relations
\begin{equation}
  \label{eq:comm-mp}
  [ \hat{A}, \hat{B} ] = i \hat{N} / N_{\rm e} , \qquad [ \hat{A}, 
  \hat{N} ] = [ \hat{B}, \hat{N} ] = 0 .
\end{equation}
The measurement outcome is then a pure addition of all single-atom
contributions. And the subsequent unitary operation will feed back the
appropriately estimated part to each atom.

Let us now consider the action of the feedback loop on the quantum
state of the many-atom system.  Let $t^\mp$ be times infinitesimally
before and after the time $t$ when the feedback is applied. The action
of the feedback on the many-atom density operator $\hat{\varrho}$ then
reads
\begin{equation}
  \label{eq:loop-map}
  \hat{\MP \varrho}(t^+) = \int \! dA \, \hat{\MP U}(A) \, \hat{\MP
  M}(A \!-\! \hat{A}) \, \hat{\MP \varrho}(t^-) \, \hat{\MP
  M}^\dagger(A \!-\! \hat{A}) \,
  \hat{\MP U}^\dagger(A) .
\end{equation}
It expresses the relation between the many-atom density operators
$\hat{\MP \varrho}(t^-)$ and $\hat{\MP \varrho}(t^+)$ before and after
the feedback, respectively. The function ${\MP M}(A)$ is the
resolution-amplitude~\footnote{It is a positive operator-valued
  measure [cf. C.M.  Caves, Phys.  Rev. D {\bf 33}, 1643 (1986)].}
that describes in its operator form the measurement process with a
measurement uncertainty $\sigma$ determined by the width of
$|M(A)|^2$.  For simplicity we may take it to be a Gaussian.  The
integration signifies the classical averaging over all possible
measurement outcomes. Note that Eq.~(\ref{eq:loop-map}) can hardly be
solved due to the large number of degrees of freedom of the many-atom
density operator.

An important class of feedback mechanisms is defined by the following
action of $\hat{U}(A)$ on the observable $\hat{\MP A}$,
\begin{equation}
  \label{eq:U-trans}
  \hat{U}^\dagger(A) \, \hat{A} \, \hat{U}(A) 
  = \hat{A} + f(A) \, \hat{N} / N .
\end{equation}
It is realized by choosing for the unitary feedback operation
\begin{equation}
  \label{eq:U-exp}
  \hat{U}(A) = \exp\!\left[ - i f(A) \, \hat{B} \right] .
\end{equation}
Here the function $f(A)$ determines the response of the feedback loop
to a measured value $A$. It encodes the desired action of the feedback
and is a numerical function chosen by the experimenter.

An example may illustrate the feedback operation given in
Eq.~(\ref{eq:U-exp}): For a system containing exactly $N$ atoms
($N_{\rm e} \!=\!  N$) and choosing for example $f(A) \!=\! s (A \!+\!
A_0)$, the transformation~(\ref{eq:U-trans}) corresponds to positive
($s \!>\!  0$) or negative ($s \!<\! 0$) feedback with an offset
$A_0$. In the case of negative feedback and vanishing offset ($s \!=\!
-1$, $A_0 \!=\! 0$) we may derive from Eq.~(\ref{eq:loop-map})
relations of the single-atom quantities ${\SP A}$ and ${\SP B}$ before
and after the feedback. That is, the system is measured and fed back
as a whole, and we regard the effects on the average properties of an
individual atom in the system.

The relations for the mean values before and after the feedback read
\begin{equation}
  \big\langle \hat{\SP A} \big\rangle_{t^+} = 0 , \qquad 
  \big\langle \hat{\SP B} \big\rangle_{t^+} = \big\langle 
  \hat{\SP B} \big\rangle_{t^-} . 
\end{equation}
Thus on average the quantity ${\SP A}$ is perfectly compensated to
zero without affecting ${\SP B}$, as expected for this type of
feedback.  Conceptual differences from the behavior of a true
single-atom system are revealed when regarding the variances of these
properties.  The variance of ${\cal B}$ reflects the expected increase
of noise due to the back-action of the measurement:
\begin{equation}
  \left\langle \big(\Delta\hat{\SP B} \big)^2 \right\rangle_{t^+} = 
  \left\langle \big(\Delta\hat{\SP B} \big)^2 \right\rangle_{t^-} 
  + \left(\frac{1}{2 \sigma} \right)^2 .
\end{equation}
However, the variance of ${\SP A}$, i.e., of the single-atom version
of the measured observable, does not correspond to that expected for
the feedback acting on a single-atom system:
\begin{equation}
  \label{eq:a-var}
  \left\langle \big(\Delta\hat{\SP A} \big)^2 \right\rangle_{t^+} = 
  \left\langle \big(\Delta \hat{\SP A} \big)^2 \right\rangle_{t^-} 
  \!\!- \frac{1}{N^2} \left\langle \big(\Delta \hat{\MP A}\big)^2
  \right\rangle_{t^-} \!\!+ \left( \frac{\sigma}{N} \right)^2 . 
\end{equation}

For $N\!=\!1$ atom in the system we have $\hat{A} \!\to\! \hat{\SP A}$
and therefore the first two terms on the r.h.s. cancel each other,
leaving as variance only the measurement uncertainty $\sigma^2$. This
is the result that we expect for the single-atom case.  In the general
$N$ atom case, however, the first two terms do not cancel each other.
Equation (\ref{eq:a-var}) can be interpreted as follows: The
uncertainty of the single-atom observable ${\SP A}$ is reduced by the
gain of knowledge on the macroscopic observable $A$.  However, the
uncertainty of the single-atom quantity ${\SP A}$ is in general not
identical to the uncertainty of the macroscopic observable $\hat{A}$
divided by the number of atoms. And therefore the first two terms do
not compensate and in this way increase the noise above the
measurement uncertainty term.

The second term in Eq.~(\ref{eq:a-var}) contains atom-atom
correlations of the type $\langle \hat{\phi}_\mu^\dagger
\hat{\phi}_\lambda \hat{\phi}_{\mu'}^\dagger \hat{\phi}_{\lambda'}
\rangle$, as can be seen from the definition (\ref{eq:ops-mp}).
Therefore, the amount of compensation of noise in the single-atom
properties due to the feedback depends on many-atom quantum
correlations. The determination of these correlations requires a
hierarchy of equations for higher-order atom correlations. To
incorporate all these correlations in a consistent way is therefore
required even for obtaining single-atom properties.

\section{Solution in terms of correlation functions} \label{sec:3}

Our goal is to describe expectation values of arbitrary extensive
properties of the system, such as for example the total energy. Any
such quantity can be expressed in terms of the so-called single-atom
density matrix $\rho_{\mu\lambda}$, that is defined by
\begin{equation}
  \label{eq:def-sp-dens-mat}
  \rho_{\mu\lambda} = \langle \hat{\phi}_\lambda^\dagger
  \hat{\phi}_\mu \rangle .
\end{equation}
This density matrix describes the average properties of single atoms
in the system and is normalized to the total average number of atoms
\begin{equation}
  \sum_\lambda \rho_{\lambda\lambda} = \langle \hat{N} \rangle .
\end{equation}

However, for the purpose of including the necessary higher-order atom
correlations and for obtaining a closed set of equations we study a
slightly different quantity. Namely, we consider the change due to a
feedback operation of the correlation function,
\begin{equation}
  \label{eq:corr-def}
  {\MP D}_{\mu\lambda}(z) = \left\langle \hat{\phi}_\lambda^\dagger \,
  \hat{\MP T}(z) \, \hat{\phi}_\mu \right\rangle .
\end{equation}
The operator $\hat{\MP T}(z)$ is yet unknown, but we choose it to
satisfy the requirement
\begin{equation}
  \lim_{z\to 0} \hat{\MP T}(z) = 1 .
\end{equation}
In this way the single-atom density matrix results as the value for
$z\!=\! 0$,
\begin{equation}
  {\MP D}_{\mu\lambda}(0) =  \langle
  \hat{\phi}_\lambda^\dagger \hat{\phi}_\mu \rangle . 
\end{equation}
As already mentioned, with the introduction of $\hat{\MP T}(z)$
higher-order many-atom correlations are contained in ${\MP
  D}_{\mu\lambda}(z)$, which is the clue to the solution.

From Eq.~(\ref{eq:loop-map}) the correlation ${\MP
  D}_{\mu\lambda}(z,t^+)$ right after a feedback operation can be
expressed as
\begin{eqnarray}
  \label{eq:loop-map2}
  {\MP D}_{\mu\lambda}(z,t^+) & = & \int \! dA \, \Big\langle 
    \hat{M}^\dagger(A \!-\! \hat{A}) \, \hat{U}^\dagger(A) \, 
    \hat{\phi}_\lambda^\dagger \, \hat{T}(z) \, \hat{\phi}_\mu 
    \nonumber \\
    & & \qquad \times \, 
    \hat{U}(A) \, \hat{M}(A \!-\! \hat{A}) \Big\rangle_{t^-} .  
\end{eqnarray}
Now the operator function $\hat{T}(z)$ is chosen in such a way, that
the r.h.s.~of this equation can again be expressed in terms of the
correlation functions. More precisely, in terms of ${\MP
  D}_{\mu\lambda}(z,t^-)$ at the time $t^-$ before the feedback
operation.  It can be shown that the specific Ansatz
\begin{equation}
  \label{eq:ansatz}
  \hat{\MP T}({\mathbf z}) = \exp( i \, {\mathbf
  z}^{\rm T} \!\cdot\! \hat{\mathbf Z}) ,
\end{equation}
is appropriate for that purpose.  Here the vector parameter is
${\mathbf z}^{\rm T} \!=\! (\alpha, \beta, \gamma)$
\footnote{${\mathbf z}^{\rm T}$ means the transpose of ${\mathbf z}$.}
and $\hat{\mathbf Z}$ is defined as the vector operator
\begin{equation}
  \label{eq:def-Z}
  \hat{\mathbf
  Z} = \left( \begin{array}{c} 
    \hat{A} \\ \hat{B} \\ \hat{N}/N_{\rm e} 
    \end{array} \right) .
\end{equation}

Using this Ansatz, from Eq.~(\ref{eq:loop-map2}) we obtain a closed
set of equations that connects the correlations ${\MP
  D}_{\mu\lambda}({\mathbf z})$ immediately after the feedback
operation with those before the feedback:
\begin{equation}
  \label{eq:loop-map4}
  {\MP D}_{\mu\lambda}({\mathbf z}, t^+) = \sum_{\mu'\lambda'} 
  \int \! d^3 \!z' \, F_{\mu\lambda}^{\mu'\lambda'}({\mathbf z}, 
  {\mathbf z}') \, {\MP D}_{\mu'\lambda'}({\mathbf z}', t^-) .
\end{equation}
The function $F_{\mu\lambda}^{\mu'\lambda'}({\mathbf z}, {\mathbf
  z}')$ represents the action of the feedback mechanism and will be
denoted in the following as feedback kernel. It is defined as
\begin{widetext}
\begin{eqnarray}
  \label{eq:fb-kernel}
  \lefteqn{F_{\mu\lambda}^{\mu'\lambda'}(\alpha, \beta, \gamma, \alpha',
  \beta', \gamma') = \delta(\beta \!-\! \beta') \, \frac{1}{4\pi^2} \int
  \! dA \int \! dA' \! \int \! dA'' \! \int \! dk
  \, \delta[ \gamma \!-\! \gamma' \!+\! \alpha f(A) \!+\!
  \beta k ] } & & \\ \nonumber 
  & & \times \, 
   \exp \!\left[ i(\alpha \!-\! \alpha') (A \!-\! A') \!+\! i k A'' 
   \right] 
  \, \langle \mu | \, \hat{\SP U}(A) \, \hat{\SP M}(A' \!+\! A'' \!-\! 
  \hat{\SP
    A}) |\mu' \rangle \, 
  \langle \lambda'| \, \hat{\SP M}^\dagger(A' \!-\! A'' \!-\! 
  \hat{\SP A}) \, \hat{\SP
    U}^\dagger(A) |\lambda
  \rangle .
\end{eqnarray}
\end{widetext}
This kernel can easily be calculated since it only contains matrix
elements of the single-atom measurement operator, that is obtained
from the resolution amplitude $M(A)$ by substituting for the argument
the single-atom operator $A \!-\!  \hat{\SP A}$, i.e.,
\begin{equation}
  \label{eq:5}
  \hat{\SP M}(A \!-\! \hat{\SP A}) = M(A \!-\! B) \big|_{B =
  \hat{\SP A}} \; .
\end{equation}
Moreover, it contains the single-atom unitary operation that reads
\begin{equation}
  \label{eq:U-sp}
  \hat{\SP U}(A) = \exp[ -i f(A) \,
  \hat{\SP B} / N_{\rm e}] \, .
\end{equation}

The kernel~(\ref{eq:fb-kernel}) can be interpreted in terms of
single-atom properties as follows: First the single atom's observable
$\hat{\SP A}$ is measured with outcome $A' \!\pm\! A''$ and then this
observable is shifted by $f(A) / N_{\rm e}$, as would be expected when
equally distributing the collective shift $f(A)$ among the estimated
number of atoms. The connection between $A$ and $A' \!\pm\! A''$,
however, is determined by the mapping of the additionally introduced
degrees of freedom, ${\mathbf z}$ and ${\mathbf z}'$. This latter
mechanism is the way how all orders of many-atom correlations are
incorporated in the mapping and signifies the deviation from the true
single-atom case.

Considering the case ${\mathbf z} \!=\! 0$ ($\alpha \!=\! \beta \!=\!
\gamma \!=\!  0$), from Eqs~(\ref{eq:loop-map4}) and
(\ref{eq:fb-kernel}) it becomes immediately clear that for the
single-atom density matrix $\langle \hat{\phi}_\lambda^\dagger
\hat{\phi}_\mu \rangle$ no closed mapping is obtained:
\begin{widetext}
\begin{eqnarray}
  \label{eq:zero-loop}
  \left\langle \hat{\phi}_\lambda^\dagger \hat{\phi}_\mu
  \right\rangle_{t^+} 
  & = &  \frac{1}{2\pi} \sum_{\mu'\lambda'} \int \! d\alpha' \! \int
  \! dA \int \! dA'
  \, e^{- i \alpha' (A - A')} 
  \, \langle \mu | \, \hat{\SP U}(A) \, \hat{\SP M}(A' \!-\! \hat{\SP
    A}) |\mu' \rangle \nonumber \\
  & & \qquad \times \,
  \langle \lambda'| \, \hat{\SP M}^\dagger(A' \!-\! \hat{\SP A}) \, \hat{\SP
    U}^\dagger(A) |\lambda
  \rangle \, D_{\mu'\lambda'}(\alpha',t^-) ,
\end{eqnarray}
\end{widetext}
where 
\begin{equation}
  D_{\mu'\lambda'}(\alpha',t^-) = D_{\mu'\lambda'}({\mathbf
    z}',t^-) \big|_{{\mathbf z}^{\prime {\rm T}} = (\alpha', 0, 0)} .
\end{equation}
On the r.h.s.~of Eq.~(\ref{eq:zero-loop}) higher-order many-atom
correlations always remain in $D_{\mu'\lambda'}(\alpha',t^-)$ due to
the non-vanishing integration variable $\alpha'$.  That is, even when
considering the single-atom dynamics, the mapping of the full
correlations $D_{\mu\lambda}({\mathbf z})$ has to be employed.
Afterwards ${\mathbf z}$ can be set to zero to obtain the sought
single-atom density matrix.

It is worth mentioning, that an approximate closed mapping for the
single-atom density matrix in fact exists in the limit $\sigma \!\to\!
\infty$. Then the measurement resolution amplitude ${\SP M}$ only
slowly depends on its argument and the $A'$ integration in
(\ref{eq:zero-loop}) leads to non-vanishing contributions only for
$\alpha' \!\approx\! 0$. That shows that for very bad measurement
resolutions, almost no correlations are generated between the atoms.

In the special case when exactly one atom is present in the system,
the correlation function will naturally reduce to the single-atom
density operator $D_{\mu\lambda}({\mathbf z}) \!\to\!  \langle
\hat{\phi}_\lambda^\dagger \hat{\phi}_\mu \rangle$.  Therefore, it
does no longer depend on ${\mathbf z}$. In this true single-atom case,
the $\alpha'$ integration in the mapping (\ref{eq:zero-loop}) acts
only on the kernel and results in the usual relation of the form given
in Eq.~(\ref{eq:loop-map}). Now, however, for the single-atom density
operator of a true single-atom system. Thus, the need of the parameter
${\mathbf z}$ signifies the presence of many-atom correlations in the
feedback mapping.

Between two feedback operations the many-atom system will evolve
according to the Hamiltonian $\hat{\MP H}$.  Assuming non-interacting
atoms it is of the form
\begin{equation}
  \label{eq:free-hamilton}
  \hat{\MP H} = \sum_{\mu\lambda} \langle \mu | \hat{\SP H} | \lambda
  \rangle \, \hat{\phi}_\mu^\dagger \hat{\phi}_\lambda .
\end{equation}
Without loss of generality the mode functions $u_\lambda(x) \!=\!
\langle x | \lambda\rangle$ are chosen as energy eigenstates of the
single-atom Hamiltonian $\hat{\SP H}$ with energies $E_\lambda$.
Given the unitary time-evolution operator
\begin{equation}
  \hat{\MP U}(t) = \exp ( - i \hat{\MP H} t) ,
\end{equation}
we may employ the time-dependent version of the operator
$\hat{T}({\mathbf z})$,
\begin{equation}
  \hat{\MP T}({\mathbf z},t) = \hat{\MP U}^\dagger(t) \, \hat{\MP
    T}({\mathbf z}) \, \hat{\MP U}(t) 
  = \exp\!\left[ i \, {\mathbf z}^{\rm T} \!\cdot\! \hat{\mathbf Z}(t)
    \right]  , 
\end{equation}
where $\hat{\mathbf Z}(t)$ reads
\begin{equation}
  \hat{\mathbf Z}(t) = \hat{U}^\dagger(t) \, \hat{\mathbf Z}
  \, \hat{U}(t) . 
\end{equation}
Using these definitions we arrive at the expression for the
correlation ${\MP D}_{\mu\lambda}({\mathbf z})$ propagated according
to the Hamiltonian~(\ref{eq:free-hamilton}) from time $t'$ to $t$,
\begin{equation}
  \label{eq:time-evol2}
  {\MP D}_{\mu\lambda}({\mathbf z},t) = \left\langle
    \hat{\phi}_\lambda^\dagger  
    \, \hat{\MP T}({\mathbf z}, t \!-\! t') \, \hat{\phi}_\mu
  \right\rangle_{t'}  
  \, e^{- i (E_\mu - E_\lambda) (t-t')} .
\end{equation}

Now an assumption has to be made with respect to the free time
evolution of the operators contained in $\hat{\mathbf Z}$.  We assume
that $\hat{\mathbf Z}(t)$ can be given as
\begin{equation}
  \label{eq:free-assumption}
  \hat{\mathbf Z}(t) = {\mathbf V}_z(t \!-\! t') \cdot \hat{\mathbf
  Z}(t') ,
\end{equation}
where the $3 \!\times\! 3$ real-valued matrix has the property
${\mathbf V}_z(-t) \!=\! {\mathbf V}^{-1}_z(t)$.  Note that the free
time evolution of many physical systems can be cast into this form.
Representative examples are free or harmonically trapped atoms when
$\hat{\MP A}$ and $\hat{\MP B}$ are center-of-mass position and total
momentum or vice versa.  In such cases the correlations evolve as
\begin{equation}
  \label{eq:free-map}
  {\MP D}_{\mu\lambda}({\bf z}, t) = \int \! d^3\!z' 
  \, W_{\mu\lambda}({\mathbf z}, {\mathbf z}', t \!-\!
  t') \, {\MP D}_{\mu\lambda}({\mathbf z'}, t') ,
\end{equation}
where in energy representation the kernel is given as
\begin{equation}
  \label{eq:free-kernel}
  W_{\mu\lambda}({\mathbf z}, {\mathbf z}', t ) = \delta^{(3)} 
  \!\left[ {\mathbf
      z'} \!-\! {\mathbf V}_z^{\rm T}(t) \!\cdot\! {\mathbf z} \right] \,
  e^{-i (E_\mu - E_\lambda) t} ,
\end{equation}
which again can be evaluated rather easily.

Now it becomes also clear, why the additional parameter $\beta$ has
been introduced. Clearly it is not necessary for describing the effect
of a feedback operation on the single-atom density matrix, as can be
seen from Eq.~(\ref{eq:zero-loop}). For the free evolution, however,
this parameter is needed to provide a complete set of dynamically
evolving variables.  Consider, for example, the case of harmonically
trapped atoms where the feedback is performed on the total momentum.
Then $\alpha$ is associated with the total momentum, whereas $\beta$
is associated with the center-of-mass coordinate. The two parameters
are then coupled during the free evolution, which shows the need for
the additional parameter $\beta$. Note, that the parameter $\gamma$
naturally emerged due to the commutator relation of $\hat{A}$ and
$\hat{B}$.

It has been shown that by introduction of an additional (real-valued)
vector parameter ${\mathbf z}$ a specifically chosen correlation
function can be evolved in time by closed sets of equations of motion.
Both for the feedback of a macroscopic observable and for the free
evolution between the feedback operations [under the
constraint~(\ref{eq:free-assumption})] the appropriate kernel
functions can be given.  The vector ${\mathbf z}$ could be formally
interpreted as being an additional degree of freedom of the single
atom, so that we may view the complete feedback dynamics as a quasi
single-atom problem.  This is justified also by the fact that all
required kernel functions are obtained from matrix elements of
single-atom operators.

The correlation function ${\MP D}_{\mu\lambda}({\mathbf z})$ reveals
another interesting aspect. Clearly, for ${\mathbf z} \!=\! 0$ it
converges to the single-atom density matrix $\rho_{\mu\lambda}$ and
describes the microscopic properties of the average single atom. It
has however also a connection to macroscopic properties, as will be
shown now. If we would neglect in the definition~(\ref{eq:corr-def})
the field operators $\hat{\phi}_\lambda^\dagger$ and $\hat{\phi}_\mu$,
we could approximately write
\begin{equation}
  D_{\mu\lambda}({\mathbf z}) \sim
  D({\mathbf z}) = \langle \hat{T}({\mathbf z}) \rangle .
\end{equation}
Given the definition of $\hat{T}({\mathbf z})$ in
Eq.~(\ref{eq:ansatz}) it becomes clear that $D({\mathbf z})$ is the
characteristic function (i.e. Fourier transform) of the probability of
measuring simultaneously the observables $\hat{A}$, $\hat{B}$, and
$\hat{N}/N_{\rm e}$. For a system with precise atom number, the latter
observable can be omitted and we essentially arrive at the Fourier
transform of the Wigner function $W(A,B)$ of the pair of canonically
conjugate macroscopic variables $\hat{A}$ and $\hat{B}$.

Turning back to the true definition~(\ref{eq:corr-def}) it can now be
observed that the diagonal correlation $D_{\lambda\lambda}({\mathbf
  z})$ can also be interpreted as the Fourier transform of a
Wigner-type function of the macroscopic phase space.  However, it is
that of a many-atom state, where one atom in mode $\lambda$ has been
removed. The off-diagonal elements $D_{\mu\lambda}({\mathbf z})$
($\lambda \!\neq\!\mu$) represent the corresponding terms including
single-atom coherences.  Thus, the chosen correlation function covers
both microscopic and macroscopic properties. Moreover, it accomplishes
this in a specific combination, that allows for the solution of the
feedback dynamics.

\section{Application to an ideal Bose gas} \label{sec:4}

For illustrating the effects of atom correlations that can be treated
with the presented method, let us consider feedback of a
gas of $N$ trapped bosonic atoms. We assume all the atoms to be
initially in the ground state of the harmonic trap potential, e.g., a
simple model for a non-interacting Bose--Einstein condensate.  The
feedback shall consist in measuring the total momentum $\hat{\MP P}$
of all atoms and in subsequently shifting this observable to zero by
use of the observed value and the center-of-mass operator $\hat{\MP
  Q}$ [$\hat{\MP A} \!\to\!  \hat{\MP P}$, $\hat{\MP B} \!\to\!
\hat{\MP Q}$, $f(P) \!=\! -P$, and $N_{\rm e} \!=\! N$].  

The initial correlation function of the degenerate bosonic gas reads
\begin{eqnarray}
  \label{eq:ini-corr-bec}
  \lefteqn{ D_{\mu\lambda}({\bf z}) = N \, \langle \mu | 0 \rangle
    \langle 0 | \lambda \rangle \exp\!\left[ i\gamma (1 \!-\! 1/N)
    \right] }  & & \nonumber \\ 
  & & \; \times \, \exp\!\left\{ -
    \frac{N \!-\! 1}{2} \left[ ({\Delta p}_0)^2
      \alpha^2 + \frac{({\Delta q}_0)^2}{N^2} \beta^2 \right] \right\} .
\end{eqnarray}
Here ${\Delta q}_0$ and ${\Delta p}_0$ are the position and momentum
uncertainties, respectively, of the single-atom ground state
$|0\rangle$ in the trapping potential.  This example shows quite well
the above mentioned combination of microscopic and macroscopic
descriptions. In this special case the
correlation~(\ref{eq:ini-corr-bec}) is a product of the single atom
density matrix $\rho_{\mu\lambda} \!=\! N \langle \mu | 0 \rangle
\langle 0 | \lambda \rangle$, a phase factor determined by the precise
number of atoms, and a Gaussian distribution. The latter is the
Fourier transform of the joint probability distribution for
simultaneously observing the macroscopic values $Q$ and $P$ of
$N\!-\!1$ atoms. That is, it is the Fourier transform of the Wigner
function $W(Q,P)$ of the center-of-mass degree of freedom of the
macroscopic atomic gas with one atom being removed.

\begin{figure}
  \begin{center}
    \epsfig{file=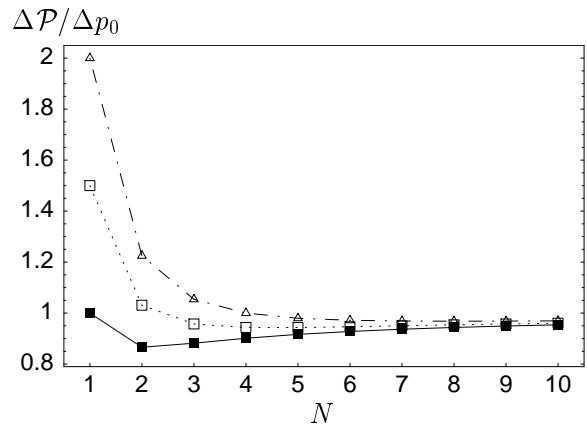,scale=0.75}
    \vspace*{-1ex}
    \caption{Variance of the single-atom momentum, scaled in units of
      the ground-state variance ${\Delta p}_0$, versus atom number $N$
      for the state after the feedback. Measurement uncertainties are
      $\sigma / {\Delta p}_0 \!=\!  1$ (filled boxes) $1.5$ (empty
      boxes), and $2$ (triangles).}
    \label{fig:example}
  \end{center}
\end{figure}

Though the correlation~(\ref{eq:ini-corr-bec}) could be propagated in
time as described previously, we consider here only the result of a
single feedback operation. In this way, the variances of single-atom
quantities, such as position and momentum, can be studied in
dependence on the atom number $N$ and the measurement uncertainty
$\sigma$.  Figure~\ref{fig:example} shows the variance of the
single-atom momentum in dependence on the atom number for three
different measurement uncertainties $\sigma$.  
One observes that for larger numbers of atoms, the momentum variance
converges to the initial ground-state variance ${\Delta p}_0$, which
establishes the regime of weak measurements where the single atom is
only weakly affected by the measurement of the feedback loop. For only
a few atoms the dependence on the chosen measurement resolution
indicates an increasing amount of atom-atom correlations with
decreasing measurement uncertainty. For larger measurement
uncertainties a $\sigma/N$ behavior will appear in the single-atom
momentum variance, as would be expected classically when the
measurement uncertainty is equally distributed among all atoms
[cf.~Eq.~(\ref{eq:a-var})].

\begin{figure}
  \begin{center}
    \epsfig{file=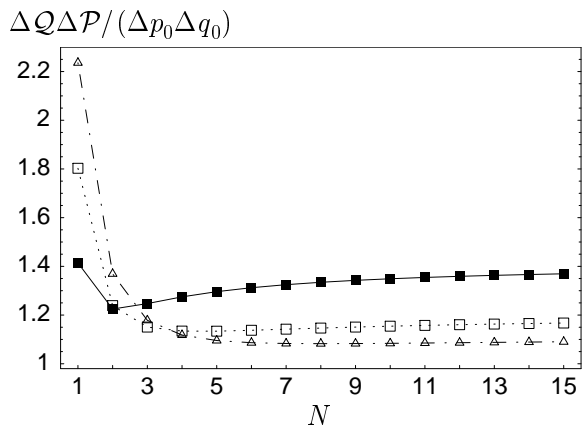,scale=0.75} \vspace*{-1ex}
    \caption{Uncertainty product of single-atom position and momentum,
      scaled in units of the ground-state uncertainty ${\Delta q}_0
      {\Delta p}_0$, versus atom number $N$ (same parameters as in
      Fig.~2). The minimum uncertainty product corresponds to the
      value $1$.}
    \label{fig:example2}
  \end{center}
\end{figure}

In Fig.~\ref{fig:example2} the uncertainty product of single-atom
momentum and position, including now also the measurement-induced
back-action noise in the position, is plotted versus the atom number
for different measurement uncertainties.  
For small atom numbers atom-atom correlations can be build up, as in
the case $\sigma / {\Delta p}_0 \!=\! 1$ (filled boxes), leading to
minimum values at certain atom numbers. For larger atom numbers the
measurement-induced noise dominates and the scaled value $[1 \!+\!
({\Delta p}_0/\sigma)^2]^{\frac{1}{2}}$ is reached.  The presented
features are those of a condensate state, for other highly correlated
quantum states one may expect even more dramatic deviations from the
classically expected behavior.

\section{Summary and conclusions} \label{sec:5}

In conclusion a method has been presented for treating the dynamics of
feedback of macroscopic observables of a bosonic multi-mode many-atom
system. The introduced unique correlation function obeys closed
equations of motion for the feedback operation and the free-evolution
dynamics. Thus the dynamics intermitted by feedback operations at
predefined times can be exactly described and calculated on a quasi
single-atom level. The description contains the full many-atom
correlations, that are generated by the feedback loop. In this way the
dynamics of expectation values of any extensive property of the
many-atom system, as for example the energy of a dilute atomic gas,
can be obtained.  The advantage of the method is its exactness
combined with the treatment on a quasi single-atom level, vastly
reducing the number of degrees of freedom, that would have to be
employed for a solution of the full many-atom problem.  Applications
of the method can be found in the context of control and stochastic
cooling of dilute bosonic gases~\cite{raizen} or feedback of a few
atoms in an optical cavity~\cite{rempe1,rempe2}.

The author thanks V.V.~Kozlov, P.E.~Toschek, I.A.~Walmsley, and
W.~Vogel for comments.  This research was supported by Deutsche
Forschungsgemeinschaft and Deutscher Akademischer Austausch Dienst.

\end{document}